# Quantitative Prediction of Linear B-Cell Epitopes

**Raul Isea**

Fundacion IDEA, Edif. Bolivar, Hoyo de la Puerta, Baruta, Venezuela

**Email address:**
risea@idea.gob.ve

**Abstract:** In scientific literature, there are many programs that predict linear B-cell epitopes from a protein sequence. Each program generates multiple B-cell epitopes that can be individually studied. This paper defines a function called <C> that combines results from five different prediction programs concerning the linear B-cell epitopes (ie., BebiPred, EPMLR, BCPred, ABCPred and Emini Prediction) for selecting the best B-cell epitopes. We obtained 17 potential linear B cells consensus epitopes from Glycoprotein E from serotype IV of the dengue virus for exploring new possibilities in vaccine development. The direct implication of the results obtained is to open the way to experimentally validate more epitopes to increase the efficiency of the available treatments against dengue and to explore the methodology in other diseases.

## 1. Introduction

Bioinformatics has contributed to the *in silico* design of vaccines by identifying immunologically significant sequences, using recently developed computational methodologies [1-3]. Given the exponential growth of biological information and technological advances in the management of data, it has been suggested that the concept of bioinformatics and computational biology should undergo a revision and use more frequently a working hypothesis containing all the available information [4, 5]. An example of this is the design of a vaccine that could be useful in malaria, based on a poly (amidoamine) (PAMAM) eight-branched dendrimeric structure formed by a set of B-cell epitopes [6].

The present paper focused on selecting the best linear B-cell epitopes obtained by BePiPred, EPMLR, BCPred, ABCPred, and Emini prediction programs according to Isea methodology [2, 3, 5, 6]. This methodology was applied on the dengue virus serotype that was detected in Venezuela for exploring new possibilities in vaccine development. It is important to consider that any proposed dengue vaccine must fulfill a series of requirements such as providing a long-term protection and a heterotypic protection. Also, the vaccine must be able to protect the individual against the four serotypes of dengue virus as well as have a low production cost.

As explained in [7-10], the first step to predict linear B-cell epitopes is to identify the protein targets, and in the case of dengue virus, the glycoprotein E is the first option to find the B epitope. In fact, glycoprotein E and the C and M proteins were tested in the immunization assays of mice and were proven to be capable of providing protection against lethal doses of dengue virus [11]. This article only considered the dengue virus E glycoprotein.

The next section will describe the methodology for the quantitative selection of potential linear B-cell epitopes that are generated by BePiPred, EPMLR, BCPred, ABCPred, and Emini Prediction programs with the sequences of dengue serotype 4 that currently exists in the Venezuelan outbreak

## 2. Materials and Methods

First, the protein sequences of the glycoprotein E in the NCBI database have to be identified. This paper only considers the serotype 4 that was isolated in Venezuela. The linear B-cell epitopes were predicted with BePiPred [12], EPMLR [13], BCPred [14], ABCPred [15], and Emini Prediction programs [16]. Given that each program generates a wide variety of potential B-cells epitopes, a strategy was developed to select the best set of epitopes. This procedure is as follows:

a. The epitopes that overlap in at least three or more regions of the prediction were selected. For example, the following epitopes were predicted: VMKRYSAPSE (region between amino acids 1 until 10), RYSAPSESE (region 4 until 12), YSAPSESEGV (5 until 14), and YSAPSESEGVL (5 until 15). Then, the four epitope predictions were aligned to overlap in the same region as illustrated below:

```
Position:  1    5    10   15
           VMKRYSAPSE
              RYSAPSESE
               YSAPSESEGV
               YSAPSESEGVL
```

In order to do this, a script in Python was developed to overlap all the B-cells epitopes in the target protein and to select the regions with three or more coincidences. According to the above example, the resulting epitope is YSAPSESE which is between the regions 5 and 12. Then, the regions where there was no consensus of at least three amino acids as shown in gray, were omitted.

b. Each amino acid was scored according to the frequency in which it appears in all predictions. This means that according to the above example, valine, methionine and lysine at positions 1, 2, and 3, respectively are equal to one. The tyrosine (position 5) until the glutamic acid (position 12) has the frequency 4, and so on.

c. The epitope obtained that overlap in three amino acids is YSAPSESE.

d. The value of <C> was determined as the sum of the frequency of amino acids divided by the total length of the predicted epitope. In the above example (i.e., YSAPSESE), <C> is (4 + 4 + 4 + 4 + 4 + 4 + 3 + 3)/8 = 3.75.

e. The epitopes whose <C> value is greater than or equal to 3.5 are selected. This restriction can be reduced to a value of 3, but it is important to select these epitopes that have a value greater than <C>.

Finally, these results are compared with the linear B-cell epitopes found in the Immune Epitope Database Analysis (IEDB) database [17]. The goal is to see if some of the B-cell epitopes that have already been used in scientific literature can be reproduced.

## 3. Results and Discussions

31 sequences of the glycoprotein E of the dengue virus serotype 4 were found in the National Center for Biotechnology Information (NCBI) database, whose ID are: ACH61691, ACH61692, ACH99663, ACH99664, ACL99023, ACL99024, ACL99026, ACL99031, ACL99035, ACO06199, ACQ44389, ACQ44390, ACQ44392, ACQ44393, ACQ44395, ACQ44396, ACQ44397, ACQ44398, ACQ44399, ACQ44400, ACQ44401, ACS32010, ACW82930, ACW82931, ACW82932, ACW82933, AEA50926, AEA50927, AET43237, AET43240, and AEX09559.

The linear B-cell of each sequence found in the NCBI database was predicted using the five programs indicated before [7-10]. Each program generates multiple B-cell epitopes as shown in the following example. Consider the E glycoprotein whose ID is ACH61691 (length 3387 aa). The BepiPred obtained 97 lineal B-cell epitopes whose length is equal to five or more amino acids (the threshold is 0.35 for this program). The ABCPred predicted 104 epitopes of the length of 10 amino acids whose score is equal to or greater than 0.7 (threshold is 0.51). The EPMLR obtained 208 epitopes of the length of 9 amino acids whose score is equal to or greater than 0.7 (threshold is -015). The program BCPred based on the flexible prediction method obtained 144 epitopes with a specificity of 75%. The Emili procedure generated 83 epitopes with a threshold of 1.0 and a window size of 3. Thus, the sequence ACH61691 yielded 636 linear B-cell epitopes.

Glycoprotein E from serotype IV of the dengue virus has 31 different sequences in the NCBI database, resulting in more than 1,800 linear epitopes. Using the above-mentioned methodology it was possible to remove the epitopes that overlap in at least three or more regions in all prediction programs. After applying the above procedure for all E glycoproteins, seventeen epitopes were obtained that have a value of <C> greater than 3.6 as indicated in table 1. According to the results shown in table 1, only seven epitopes have a value of <C> greater than four.

*Table 1.* Linear B-cell epitopes generated for all glycoprotein E of dengue serotype 4 present in Venezuela with <C> value equal to or greater than 3.6 (see text for details).

| Region | Epitope | <C> |
| --- | --- | --- |
| 909-924 | IDGPDTSECPNERRAW | 4.25 |
| 1811-1820 | REIPERSWNT | 4.20 |
| 1983-1992 | FGPEREKTQA | 4.20 |
| 807-817 | KFQPESPARLA | 4.10 |
| 1643-1653 | ERIGEPDYEVD | 4.10 |
| 2056-2066 | WTREGEKKKLR | 4.10 |
| 2040-2047 | GERNNQIL | 4.00 |
| 2848-2856 | VDTRTPQPKPG | 3.82 |
| 1438-1545 | RLEPSWADV | 3.78 |
| 1790-1805 | ATPPGATDPFPQSNSP | 3.75 |
| 2126-2133 | TERGGRAY | 3.75 |
| 3179-3197 | IPQWEPSKGWKNWQEVPFC | 3.74 |
| 2214-2222 | EPEKQRTPQ | 3.67 |
| 196-203 | SGERRREK | 3.63 |
| 1938-1945 | NPAQEDDQ | 3.63 |
| 1556-1568 | GWRLGDKWDKEED | 3.62 |
| 647-655 | IELEPPFGDS | 3.60 |

Finally, the entire IEDB database was searched for all the lineal B-cells epitopes used in the experiments based on the dengue serotype 4. The result is shown in table 2. When comparing the two tables, it is evident that there are coincidences between the computational methodology presented in this paper and that found in IEDB, which are highlighted in bold type. These coincidences indicate that it is possible to reduce the number of linear B cells that can be used in an experiment.

## 4. Conclusions

As explained above, one protein sequence can generate multiple linear B-cell epitopes and for this reason, this paper proposes a computational methodology to reduce the number of linear B-cell epitopes and thus allow one to rationally select those B-cell epitopes that can be predicted more accurately.

By comparing the results obtained in table 1 with respect to the epitopes obtained from the scientific literature given in table 2, it can be seen that there is a correlation between the two. However, it is necessary to verify these results with experimental tests that will be conducted in the future. Once there is agreement between the experiment and this methodology, a web version will be developed using the python script.

*Table 2. Linear B-cell epitopes obtained from the IEDB database corresponding to dengue serotype 4 data.*

| ID IEDB | Epitope |
|---|---|
| 16649 | FL**IDGPDTSECPNERRA** |
| 27016 | ILEENMEVEI**WTREGEKKKL** |
| 30831 | **KFQPESPARLA**SAILNA |
| 41821 | MKFREGSSEVC |
| 73272 | WYGMEIRPLSEKEENMV |
| 145953 | PWHLGKLEIDFGECPGTTVTIQEDCDHRGPSLRTTTASGKLVTQWC |
| 145998 | **SECPNERRAW**NFLEVEDYGFGMFTTNIW |
| 151601 | ADLSLEKAANVQWDE |
| 179894 | RTTTASGKLVT |

## Acknowledgements

The author wishes to express his sincere thanks to Prof. Silvia Restrepo and John Hoebeke for their unconditional help and the comments concerning this manuscript.